\documentclass[english,english,pra,english,preprint,amsmath,amssymb,aps,longbibliography,showkeys,titlepage]{revtex4-2}
\usepackage{lmodern}
\usepackage{lmodern}
\usepackage[T1]{fontenc}
\usepackage[utf8]{inputenc}
\usepackage{amsmath}
\usepackage{amsthm}
\usepackage{amssymb}
\usepackage{graphicx}
\usepackage{esint}

\makeatletter
\usepackage{amsthm}\usepackage{latexsym}\usepackage{bm}\usepackage{amsfonts}
\usepackage{babel}
\usepackage{hyperref}
\usepackage{enumitem}

\usepackage{hyperref}
\hypersetup{
	breaklinks = true,
    colorlinks = true,
    citecolor = {blue},
	urlcolor = {blue},
	linkcolor = {blue}
}

\makeatother

\usepackage{babel}
\begin{document}
\title{Nonlinear solution of classical three-wave interaction via finite
dimensional quantum model}
\author{Michael Q. May}
\email{mqmay@princeton.edu}

\affiliation{Plasma Physics Laboratory, Princeton University, Princeton, NJ 08540,
U.S.A}
\author{Hong Qin}
\email{hongqin@princeton.edu}

\affiliation{Plasma Physics Laboratory, Princeton University, Princeton, NJ 08540,
U.S.A}
\begin{abstract}
The quantum three-wave interaction, the lowest order nonlinear interaction
in plasma physics, describes energy-momentum transfer between three
resonant waves in the quantum regime. We describe how it may also
act as a finite-degree-of-freedom approximation to the classical three-wave
interaction in certain circumstances. By promoting the field variables
to operators, we quantize the classical system, show that the quantum
system has more free parameters than the classical system, and explain
how these parameters may be selected to optimize either initial or
long-term correspondence. We then numerically compare the long-time
quantum/classical correspondence far from the fixed point dynamics.
We discuss the Poincaré recurrence of the system and the mitigation
of quantum scrambling. 
\end{abstract}
\maketitle

\section{Introduction}

The three-wave interaction equations may describe the dynamics of
the nonlinear interactions of three waves, or they can describe individual
or pairs of small-amplitude waves in nonlinear media. For example,
in the decay interaction, a large amplitude wave with frequency $\omega_{1}$
and wavenumber $k_{1}$ will decay into two smaller waves with energies
$\omega_{2},\omega_{3}$ and wavenumbers $k_{2},k_{3}$ if the resonance
conditions $\omega_{1}=\omega_{2}+\omega_{3}$ and $k_{1}=k_{2}+k_{3}$
are met. The three-wave interaction equations have applications in
laser-plasma interactions \citep{moody,myatt}, determining weak turbulence
spectra \citep{zak2}, nonlinear optical system design \citep{frantz,ahn,brunton},
and oceanic wave theory \citep{usama13}. Although the three-wave
equations are well-studied \citep{rosenbluth,zakharov,Kaup1979,Reiman1979}
and their solutions (in terms of Jacobi elliptic functions \citep{armstrong})
are known, the three-wave interaction equations provide a model for
understanding nonlinear interactions in general, as they are the lowest-order
nonlinear interactions found in many systems, including plasma physics. 

Very little can be said in general about nonlinear dynamical systems,
so many techniques have been developed to transform nonlinear systems
into linear systems with a finite state-space. The transformation
to a finite, linear system also facilitates the use of quantum computers
since quantum computers act on finite-dimensional spaces with linear
operators. A popular technique is the process of Koopman embedding,
initially developed by Koopman \citep{koopman31} and Von Neumann
\citep{neumann1932} (KvN), in which the a nonlinear system is transformed
into a (potentially) infinite dimensional linear system using operational
dynamics. This operational dynamic method allows for extensions of
the Hartman-Grobman theorem, which shows that linearized dynamics
around fixed points will be qualitatively similar to the actual dynamics,
to entire basins of attraction \citep{lan2013}, and there may be
significant computational advantages to evaluating classical dynamics
rendered finite dimensional by the KvN method on a quantum computer
\citep{joseph20}.

Operational dynamic methods have several significant drawbacks, however,
which restrict their theoretical applications. For actual computation,
the linear infinite dimensional systems must be rendered finite, either
by finding a Koopman invariant subspace or using a closure. There
is research into determining Koopman-invariant finite subspaces \citep{brunton2016,mezic99,Budisic12},
but solutions are often narrowly tailored to specific systems. Operational
dynamical methods also suffer from the issue of \emph{ad hoc }linearization,
with an infinite number of Koopman linearizations possible for most
nonlinear systems. Although the original work of Koopman prescribed
a unitary embedding, this has often been ignored in Koopman-derived
research, particularly dynamical mode decomposition. Care must be
taken to ensure that the linearization and closure chosen result in
unitary linear dynamics for applications in quantum computation. 

Instead of using operational dynamics to linearize the quantum three-wave
interaction, we propose transforming the classical interaction into
a quantum interaction via the quantum field theoretic method, originally
explored for the three-wave interaction by Shi \textit{et al.} \citep{Shi2017,Shi2018,Shi2021}
and robustly simulated on a quantum computer \citep{shiPhysRevA}.
In this method, the classical dynamic variables, which represent the
waves' amplitudes, are promoted to linear operators which obey canonical
commutation relations and act on a Hilbert space. If the resultant
system is infinite dimensional, as one would expect from a Koopman
linearization, nothing would be gained from the quantization; however,
it happens that for the three-wave interaction, the quantization allows
for a natively finite-dimensional description of the dynamics. Of
course, this comes at the cost of the linear quantum system not necessarily
capturing the classical nonlinear dynamics. The quantum wave-function
will be not be localized, may be able to explore classically forbidden
regions, and will exhibit interference due to complex phase interactions.
Despite these drawbacks, elsewhere, quantum versions of classical
equations have been used to calculate classical dynamical quantities,
including diffusion coefficients and Lyaponov exponents, more efficiently
than the classical equations could \citep{joseph20,benenti01,benenti03}. 

In the following, we will use the quantum three-wave interaction as
model for the classical three-wave interaction, finding that the quantum
system is able to reproduce the classical nonlinear periodic solution
for finite times. Quantum-classical correspondence for the three-wave
interaction has been previously explored in the context of quantum
instabilities in non-chaotic classical systems \citep{may_qin}. The
nonlinear development of the quantum three-wave interaction, though,
has not been previously characterized. Through its relationship with
the classical three-wave interaction, it may serve as a fundamental
model for the application of quantum field theoretic quantization
as a means of probing other nonlinear classical symplectic systems
using linear unitary theory. 

We begin in Section \ref{sec:Three-Wave-Interactions} by reviewing
the dynamics of the classical three-wave equations. We also review
the quantum three-wave equations and their finite state-space representation.
We find that the governing equations of the classical and quantum
interactions are structurally similar and discuss their differences.
In Section \ref{sec:Comparisons} we describe the initial conditions
we will use for comparing the quantum and classical systems. We compare
the integrated classical system with the finite, linear quantum system,
finding excellent correspondence for many classical nonlinear periods.
We also explain how hyperparameters available in the quantum system,
including the initial variance and dimension, can be used to extend
the correspondence between the quantum and classical systems, even
for large nonlinearities. Finally in Section \ref{sec:Conclusions},
we summarize our findings and discuss their relevance to quantum computation. 

\section{Three-Wave Interactions\label{sec:Three-Wave-Interactions}}

The homogeneous, classical three-wave equations for the decay interaction
are given by
\begin{align}
\partial_{t}A_{1} & =gA_{2}A_{3},\label{eqn:a1}\\
\partial_{t}A_{2} & =-g^{*}A_{1}A_{3}^{*},\\
\partial_{t}A_{3} & =-g^{*}A_{1}A_{2}^{*},\label{eqn:a3}
\end{align}
where $A_{j}$ is the amplitude of the $j$-th wave, $A_{j}^{*}$
is its complex conjugate, and $g$ is the coupling coefficient \citep{jurkus,jaynes,Kaup1979,Reiman1979}.
In addition to the Hamiltonian 
\begin{equation}
H=gA_{1}^{*}A_{2}A_{3}-g^{*}A_{1}A_{2}^{*}A_{3}^{*}\thinspace,\label{eq:H_classical}
\end{equation}
the interaction supports two other constants of motion
\begin{align}
s_{2} & =I_{1}+I_{3},\label{eq:s2}\\
s_{3} & =I_{1}+I_{2},\label{eq:s3}
\end{align}
where $I_{j}=A_{j}^{*}A_{j}$ is the wave action of the $j$-th wave.
Because the wave amplitudes $A_{j}$ are not real valued, their quantum
analogs will not be observables. Thus, to directly compare the classical
and quantum three-wave interactions, we will consider the second order
differential equation for the first wave action
\begin{equation}
\partial_{t}^{2}I_{1}=\ 2|g|^{2}\left(s_{2}s_{3}-2\left(s_{2}+s_{3}\right)I_{1}+3I_{1}^{2}\right),\label{eq:classical_eqn}
\end{equation}
which is obtained directly by differentiating $I_{1}$ and making
substitutions using Eqs. (\ref{eqn:a1}---\ref{eqn:a3}) and Eqs.
(\ref{eq:s2}) and (\ref{eq:s3}). The dynamics for $I_{2}$ and $I_{3}$
are the same as those for $I_{1}$ thanks to the constants $s_{2}$
and $s_{3}$. Note that the the second order differential equation
for $I_{1}$ is decoupled from the equations for $I_{2}$ and $I_{3}$
(which is not the case for the first order differential equation for
$I_{1}$). Because there is a symmetry between $s_{2}$ and $s_{3}$,
in what follows we will assume $s_{3}\ge s_{2}$. 

We can further simplify this equation by scaling it by $s_{2}$ and
making the time parameter dimensionless so
\begin{equation}
\partial_{\tau}^{2}x=2\left(c-2(1+c)x+3x^{2}\right),\label{eq:scaled_classical_eqn}
\end{equation}
the scaled wave amplitude $x=I_{1}/s_{2},$ the constant $c=s_{3}/s_{2}$,
and the time parameter $\tau=|g|\sqrt{s_{2}}t$. In this form, the
three initial conditions $A_{1}(t=0)$, $A_{2}(t=0)$, and $A_{3}(t=0)$
necessary to integrate Eqs. (\ref{eqn:a1}---\ref{eqn:a3}) become
initial conditions on the first wave's scaled action $x$, $x(\tau=0)$
and $\partial_{\tau}x(\tau=0)$, and the constant $c$. Eq. (\ref{eq:scaled_classical_eqn})
is integrable, and one solution can be written as a Jacobi elliptic
function,
\begin{equation}
x(\tau)=c\ \text{sn}^{2}(p+\tau,c),\label{eq:jacobi}
\end{equation}
where $p$ is a constant determined by the initial conditions. Note
that although Eq. (\ref{eq:scaled_classical_eqn}) is a second order
equation, the above solution only has a single free parameter. This
is because the solution space for Eq. (\ref{eq:scaled_classical_eqn})
is much larger than that of the original problem given by Eqs. (\ref{eqn:a1}---\ref{eqn:a3}),
to which Eq. (\ref{eq:jacobi}) is also a solution. The space of initial
conditions $A_{1}(0)$, $A_{2}(0)$, and $A_{3}(0)$ is not surjective
onto the space of $c$, $x(0)$, and $\partial_{\tau}x(0)$. Indeed,
because $s_{2}$, a constant determined by the initial conditions,
has been scaled out of Eq. (\ref{eq:scaled_classical_eqn}), we are
not free to determine $x(0)$ and $\partial_{\tau}x(0)$ independently.
There are additional requirements that $c\ge1$ and $0\le x(0)\le1$.
It is also clear from Eq. (\ref{eq:scaled_classical_eqn}) that the
constant $c$ will act as the modulator of the nonlinearity of the
interaction. Taking $c=1$, its lowest value since we've assumed $s_{3}\ge s_{2}$,
maximizes the effect of the nonlinear term $x^{2}$, while taking
$c\rightarrow\infty$ the system becomes linear and the Hamiltonian
becomes that of the algebraic discrete quantum harmonic oscillator
\citep{adqo}. 

Using the quantum field theoretic method of quantization, we can promote
the wave amplitudes of Eqs. (\ref{eqn:a1}---\ref{eqn:a3}) to operators
and replace the complex conjugation with Hermitian conjugation to
obtain a set of quantum three-wave equations: 

\begin{align}
\partial_{t}\hat{A}_{1} & =g\hat{A}_{2}\hat{A}_{3},\nonumber \\
\partial_{t}\hat{A}_{2} & =-g^{*}\hat{A}_{1}\hat{A}_{3}^{\dagger},\nonumber \\
\partial_{t}\hat{A}_{3} & =-g^{*}\hat{A}_{1}\hat{A}_{2}^{\dagger}.\label{eqn:heisen}
\end{align}
The operators have the canonical commutation relations $[\hat{A}_{i},\hat{A}_{j}^{\dagger}]=\delta_{i,j}$
for $i,j\in\{1,2,3\}$ and with $\delta_{i,j}$ the Kronecker delta
function. The quantum Hamiltonian 
\begin{align}
\hat{H}=ig\hat{A}_{1}^{\dagger}\hat{A}_{2}\hat{A}_{3}-ig^{*}\hat{A}_{1}\hat{A}_{2}^{\dagger}\hat{A}_{3}^{\dagger},\label{eqn:ham}
\end{align}
and the mutually commuting operators

\begin{align}
\hat{s}_{2} & =\hat{n}_{1}+\hat{n}_{3},\label{eq:s2_quant}\\
\hat{s}_{3} & =\hat{n}_{1}+\hat{n}_{2},\label{eq:s3_quant}
\end{align}
where the number operators are defined in the usual way, $\hat{n}_{i}=\hat{A}_{i}^{\dagger}\hat{A_{i}}$.
We will denote the eigenvalues of these Hermitian operators with the
same symbols as the classical operators, e.g. $\langle\hat{s}_{2}\rangle=s_{2}$.
As found by Yuan Shi \citep{shiPhysRevA,Shi2021}, eigenvectors of
the operators $\hat{s}_{2}$ and $\hat{s}_{3}$, and the eigenvectors
of the number operators $\hat{n}_{1}$, $\hat{n}_{2}$, and $\hat{n}_{3}$
form a finite $d=s_{2}+1$ dimensional invariant subspace when acted
on by the Hamiltonian. We can write such subspace elements as
\begin{equation}
\Psi(t)=\sum_{i=0}^{s_{2}}\alpha_{i}(t)\psi_{i},\label{eqn:psi}
\end{equation}
where,
\begin{equation}
\psi_{i}=\left|s_{2}-i,s_{3}-s_{2}+i,i\right\rangle \label{eqn:psi2}
\end{equation}
has eigenvalues $\{\hat{n}_{1},\hat{n}_{2},\hat{n}_{3}\}\psi_{i}=\{s_{2}-i,s_{3}-s_{2}+i,i\}\psi_{i}$.
The finiteness of the subspace may be directly seen through the action
of the Hamiltonian on a subspace element $\psi_{i}$:

\begin{eqnarray}
H\psi_{i} & = & ig(s_{2}-i+1)^{1/2}(s_{3}-s_{2}+i)i^{1/2}\psi_{i-1}\nonumber \\
 &  & -ig^{*}(s_{2}-i)^{1/2}(s_{3}-s_{2}+1+i)(i+1)^{1/2}\psi_{i+1}.
\end{eqnarray}
If $i=s_{2}$ in the above equation, then the coefficient of $\psi_{s_{2}+1}$
will be zero and similarly for $i=0$ and $\psi_{-1}$. The action
of the Hamiltonian on these subspace elements can be calculated directly
from the Schrödinger equation
\begin{equation}
i\partial_{\tau}\Psi=H\Psi,\label{eq:schr}
\end{equation}
where we've taken the constant $\hbar=1$. Writing $\Psi(t)$ as a
column vector of weights $\left(\alpha_{0}(t),\alpha_{1}(t),\dots\alpha_{s_{2}}(t)\right)$,
$H$ becomes a $d\times d$ tridiagonal matrix,
\begin{eqnarray}
H=\left(\begin{array}{cccccc}
0 & h_{0} & 0 & 0 & 0 & \dots\\
h_{0} & 0 & h_{1} & 0 & 0 & \dots\\
0 & h_{1} & 0 & h_{2} & 0 & \dots\\
0 & 0 & h_{2} & 0 & h_{3} & \dots\\
\vdots & \vdots & \vdots & \vdots & \vdots & \ddots
\end{array}\right),
\end{eqnarray}
with 
\begin{equation}
h_{i}=\sqrt{(s_{2}-i)(s_{3}-s_{2}+1+i)(i+1)}.
\end{equation}
Thus, explicitly we may write the Schrödinger equation as a system
of $d$ coupled first-order linear differential equations: 
\begin{align}
i\dot{\alpha}_{0} & =h_{0}\alpha_{1},\label{eqn:w1-1}\\
i\dot{\alpha}_{1} & =h_{0}\alpha_{0}+h_{1}\alpha_{2},\\
 & \ \ \ \ \ \dots\nonumber \\
i\dot{\alpha}_{i} & =h_{i-1}\alpha_{i-1}+h_{i}\alpha_{i+1}.\label{eqn:w2-1}
\end{align}
We have taken $g=-i$ in the above equations for simplicity; however,
it will be shown that the phase and magnitude of $g$ will not affect
the quantum dynamics, as they did not affect the classical scaled
dynamics of Eq. (\ref{eq:scaled_classical_eqn}), below.

To compare this finite linear system with the nonlinear classical
system, we need only take the expectation value of $\hat{n}_{1}$,
\begin{equation}
\langle\hat{n}_{1}\rangle=\sum_{i=0}^{s_{2}}|\alpha_{i}|^{2}(s_{2}-i)\label{eq:n1_expectation}
\end{equation}
and compare it with the classical wave action $I_{1}.$ Of course,
the quantum linear dynamical equations and the classical nonlinear
equations refer to different systems, so their dynamics will diverge
except in the classical limit ($s_{2}\rightarrow\infty)$. Let us
find the quantum equivalent of the classical Eq. (\ref{eq:scaled_classical_eqn})
to see how the quantum dynamics might be written as a nonlinear differential
equation. First, we find a second order equation for $\hat{n}_{1}$
by using the quantum three-wave equations and making substitutions
with the definitions of $\hat{s}_{2}$ and $\hat{s}_{3}$:
\begin{equation}
\partial_{t}^{2}\hat{n}_{1}=\ 2|g|^{2}\left(\hat{s}_{2}\hat{s}_{3}-\left(2\hat{s}_{2}+2\hat{s}_{3}+1\right)\hat{n}_{1}+3\hat{n}_{1}^{2}\right).
\end{equation}
This is similar to that for $I_{1}$ in Eq. (\ref{eq:classical_eqn}).
Next, we take the expectation of this equation
\begin{equation}
\partial_{t}^{2}\langle n_{1}\rangle=\ 2|g|^{2}\left(s_{2}s_{3}-\left(2s_{2}+2s_{3}+1\right)\langle n_{1}\rangle+3\langle n_{1}\rangle^{2}+3\delta\right),
\end{equation}
where we have defined the variance 
\begin{equation}
\delta=\langle n_{1}^{2}\rangle-\langle n_{1}\rangle^{2}.
\end{equation}
Finally, we scale the equation for $\langle\hat{n}_{1}\rangle$ by
$s_{2}^{2}$, make the equation dimensionless by using the time parameter
to absorb the coupling coefficient $g$, and define $x_{Q}=\langle n_{1}\rangle/s_{2},$
$\tau=|g|\sqrt{s_{2}}t$, and $\delta^{\prime}=\delta/s_{2}^{2}$
to arrive at

\begin{equation}
\partial_{\tau}^{2}x_{Q}=2\left(c-2(1+c+\frac{1}{2s_{2}})x_{Q}+3x_{Q}^{2}+3\delta^{\prime}\right),\label{eq:scaled_quantum}
\end{equation}
which may be directly compared with Eq. (\ref{eq:scaled_classical_eqn}).
Both the quantum equation for $x_{Q}$ and the classical Eq. (\ref{eq:scaled_classical_eqn})
for $x$ are the same except for the additional linear factor of $2x_{Q}/s_{2}$
and the inclusion of the scaled variance in the quantum system. As
the dimension of the quantum system increases, so will $s_{2}$, diminishing
the effect of the $2x_{Q}/s_{2}$ term; however, the effect of the
variance in Eq. (\ref{eq:scaled_quantum}) will depend on the initial
conditions of the quantum system, not just the the total dimension.
Also, note the variance is not a function of $x_{Q}.$ It must be
calculated directly from the Schrödinger equation. 

\section{Quantum-Classical Correspondence\label{sec:Comparisons}}

The initial conditions of the quantum system must be chosen carefully
to correspond to those in the classical system. Consider an arbitrary
initial condition where we write the weights of Eq. (\ref{eq:n1_expectation})
in polar form,
\begin{equation}
\alpha_{i}=r_{i}e^{i\phi_{i}},\label{eq:n1_init_con}
\end{equation}
with a real amplitude $r_{i}$ and argument $\phi_{i}$. Using Eqs.
(\ref{eqn:w1-1}---\ref{eqn:w2-1}), we may directly differentiate
Eq. (\ref{eq:n1_expectation}) to find

\begin{equation}
\partial_{t}\langle\hat{n}_{1}(0)\rangle=2\sum_{i=0}^{s_{2}}h_{i}r_{n}r_{n+1}\text{sin}(\phi_{i+1}-\phi_{i}).\label{eq:n1_init_vel}
\end{equation}
In choosing the initial condition for the classical system, for exact
correspondence we would have $x(0)=x_{Q}(0)$, $\partial_{\tau}x(0)=\partial_{\tau}x_{Q}(0)$,
and the nonlinearity parameters equal. This system of equations is
underdetermined though, because the quantum system has many more degrees
of freedom, $f\sim O(s_{2})$, than the classical system. To deal
with this, we will restrict ourselves to considering quantum initial
conditions for which the real amplitude $r_{i}$ is taken to be a
Gaussian over the index of $\alpha_{i}$ with a mean $\mu$ and standard
deviation $\sigma$ which will depend on the dimension of the quantum
system:
\begin{equation}
r_{i}=\mathcal{N}e^{-\frac{(i-\mu s_{2})^{2}}{2\sigma^{2}}}.
\end{equation}
The normalization $\mathcal{N}$ must account for the initial condition
being clipped since $r_{-1}=r_{s_{2}+1}=0$. Note that the initial
scaled variance $\delta_{0}^{\prime}$ of the quantum system is only
qualitatively related to the standard deviation $\sigma$ since 
\begin{equation}
\delta_{0}^{\prime}=\left(\sum_{i=0}^{s_{2}}r_{i}^{2}(1-i/s_{2})^{2}\right)-\left(\sum_{i=0}^{s_{2}}r_{i}^{2}(1-i/s_{2})\right)^{2},
\end{equation}
while 
\begin{equation}
\sigma=\left(\sum_{i=0}^{s_{2}}r_{i}^{2}(i-\mu)^{2}-\left(\sum_{i=0}^{s_{2}}r_{i}^{2}(i-\mu)\right)^{2}\right)^{1/2}.
\end{equation}
Let us further restrict our quantum initial conditions to those which
are velocity maximizing, which amounts to taking $\phi_{i+1}-\phi_{i}=\pi/2$
for all $i$. This is a prescription of the initial phase of the quantum
nonlinear orbit. Since we are principally interested in correspondence
over the course of many nonlinear orbits, the initial phase should
not matter for our analysis. We thus need to match $\mu$ and $\sigma$
for the point on the classical orbit where the velocity is maximized.
As noted in Section \ref{sec:Three-Wave-Interactions}, $x(0)$ and
$\partial_{\tau}x(0)$ are not independent, so choosing to maximize
the classical velocity also specifies a point on the classical trajectory.
The solution to Eq. (\ref{eq:scaled_classical_eqn}) at the classical
inflection point (when $\partial_{\tau}^{2}x=0$) is 
\begin{equation}
x_{0}=\frac{1}{3}\left(1+c-\sqrt{1+c^{2}-c}\right).
\end{equation}
The negative root is chosen because the velocity at the inflection
point
\begin{equation}
\dot{x}_{0}=2\sqrt{x_{0}}\sqrt{1-x_{0}}\sqrt{c-x_{0}}
\end{equation}
becomes imaginary for the positive root. Finally, we set $\mu=x_{0}$.
The choice of initial standard deviation $\sigma$ will be discussed
below. 

\begin{figure}
\begin{minipage}[t]{0.45\columnwidth}%
\includegraphics[width=1.2\linewidth]{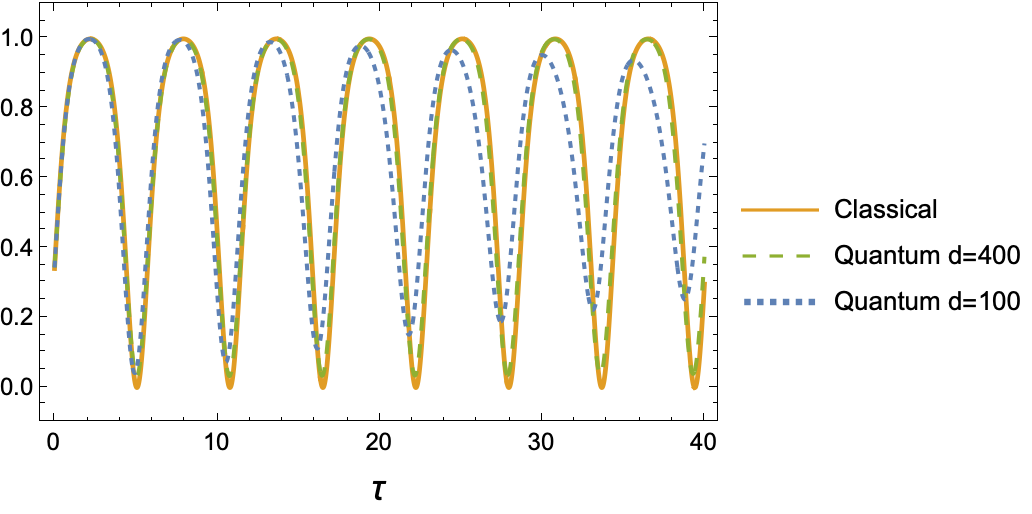}

(a) %
\end{minipage}

\caption{Comparison between nonlinear classical and quantum dynamics. The initial
standard deviation chosen for the quantum systems is set at $d/5$,
and the nonlinearity parameter $c=1.05$. }
\label{fig:quantum_classical_correspondence}
\end{figure}

\begin{figure}
\noindent\begin{minipage}[t]{1\columnwidth}%
\includegraphics[width=1\linewidth]{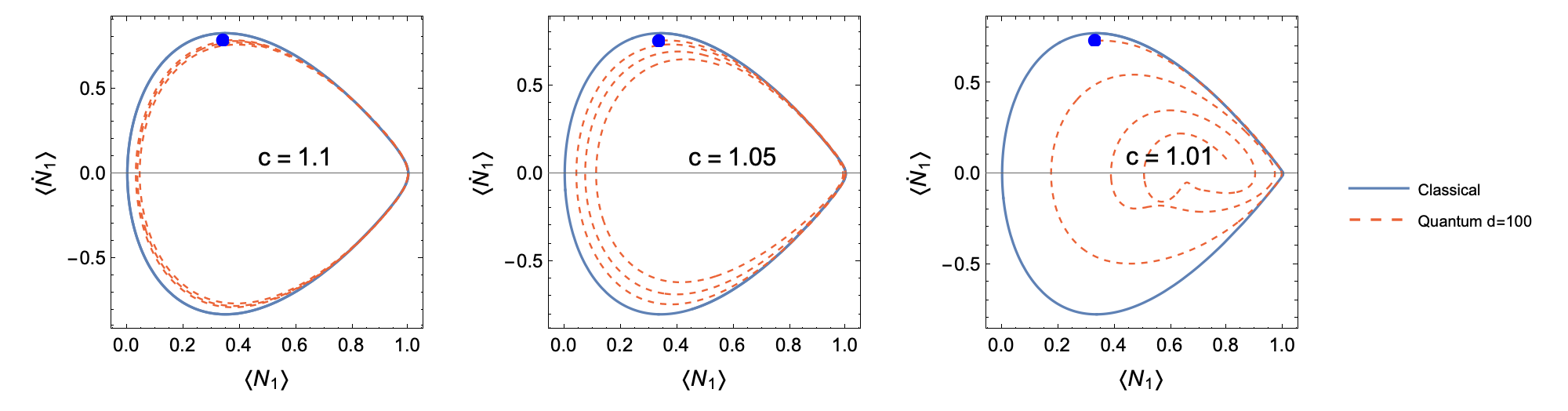}%
\end{minipage}\caption{Phase space plots of the classical $(x,\dot{x})$ and quantum $(x_{Q},\dot{x}_{Q})$
systems with initial condition $(x_{0},\dot{x}_{0})$, initial quantum
dimension $d=100$, and initial quantum standard deviation $\sigma=20$
for three values of the nonlinearity parameter $c=1.1$, $1.05$,
and $1.01$. The classical trajectories are shown in solid blue lines.
The quantum trajectories are shown via the dashed red lines. The blue
dot indicates the initial condition. Each system is evolved for three
classical orbits. \label{fig:Phase-space-plots}}
\end{figure}

We compare the integrated classical system with initial conditions
$x(0)=x_{0},$ $\partial_{\tau}x(0)=\dot{x}_{0}$ with the quantum
system in Figs. \ref{fig:quantum_classical_correspondence}, \ref{fig:Phase-space-plots},
and \ref{fig:Quantum_Revival}. As discussed above, the initial condition
is Gaussian in $r$, and we have also taken $\mu=x_{0}$, $\sigma=d/5$,
and the normalization $\mathcal{N}$ chosen such that the total probability
is 1. In what remains of this section, we will consider the quantum-classical
correspondence and various effects on that correspondence due to different
choices of nonlinearity parameter $c$, the initial standard deviation
$\sigma$, and the quantum dimension $d$. 

In Fig. \ref{fig:quantum_classical_correspondence}, we compare three
systems, two quantum with $d=100$ and $d=400$ and the classical
system, for a nonlinearity parameter $c=1.05$. Recall that $c\ge1$,
and as $c\rightarrow1$, the nonlinearity increases. Despite the quantum
and classical velocities being independently maximized, there is excellent
initial correspondence between all three systems. The nonlinearity
is pronounced enough that the $d=100$ system diverges from the classical
solution within a couple of periods; however, the correspondence for
the $d=400$ system lasts much longer, with the $d=400$ quantum systems
accurately capturing both the nonlinearity of the classical orbit
as well as its period.

We may more carefully explore the effect of the nonlinearity parameter
$c$ on the quantum-classical dynamics via phase-space diagrams of
the quantum $d=100$ and classical systems in Fig. \ref{fig:Phase-space-plots}.
Three classical nonlinear orbits are shown for each of the nonlinearity
parameters $c=1.1$, $1.05$, and $1.01$. Decreasing the nonlinearity
parameter causes the classical orbits to become more pinched. This
increasingly linear relationship between $\langle\hat{n}_{1}\rangle$
and $\partial_{\tau}\langle\hat{n}_{1}\rangle$ indicates exponential-like
growth, the onset of the classical instability and its quantum counterpart
(see \citep{may_qin}). From the quantum orbits in Fig. \ref{fig:Phase-space-plots},
it is obvious that as the nonlinearity increases, the correspondence
between the systems decreases. Investigations into exponential growth
of quantum correlators in non-chaotic systems indicate that proximity
to the classical fixed point leads to quantum scrambling \citep{xu20}.
For instance, after a single pass near the classical fixed point,
the $d=100$ system sharply diverges from the classical solution for
$c=1.01$. On the other hand, the $d=100$ quantum system approximates
the classical solution for a couple of classical periods for $c=1.05$
(see also Fig. \ref{fig:quantum_classical_correspondence}), and the
correspondence lasts for much longer for $c=1.1$. The proximate cause
of the sharp divergence of the quantum system from the classical system
near the fixed point is the increased classical period as $c\rightarrow1$.
Since the quantum system may only approximate the classical system
for finite times, when $c\rightarrow1$ and the classical period tends
to infinity, the quantum solution will inevitably diverge from the
classical within a single period. 

\begin{figure}
\includegraphics[scale=0.75]{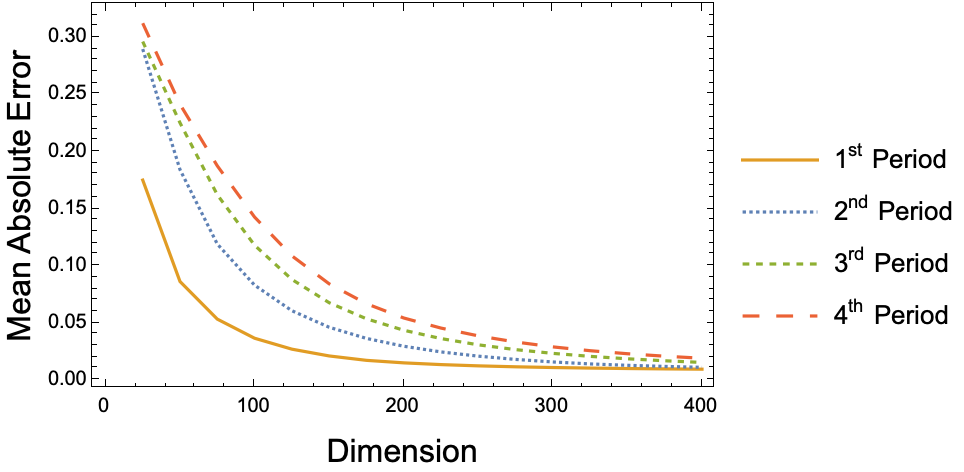}\caption{Mean absolute error error between the classical and quantum systems
with $c=1.05$, averaged over each of the first four classical periods.
The dimension $d$ ranges from 25 to 400.}
\label{fig:error}
\end{figure}

Shown in Fig. \ref{fig:error} is the mean absolute error between
the classical system and quantum systems with a range of dimensions
and nonlinearity parameter $c=1.05$. The error is averaged over one
period of the classical orbit for each of the first four classical
periods. Note that for two uncorrelated oscillators with amplitude
1, the expected value of the mean absolute error will be 0.25. A mean
absolute error higher than this must be due to anti-correlated phases.
This occurs for low dimension in Fig. \ref{fig:error}. As the dimension
of the quantum simulation is increased, the error decays quickly;
doubling the dimension of the quantum system more than halves the
mean absolute error.

\begin{figure}
\includegraphics[scale=0.75]{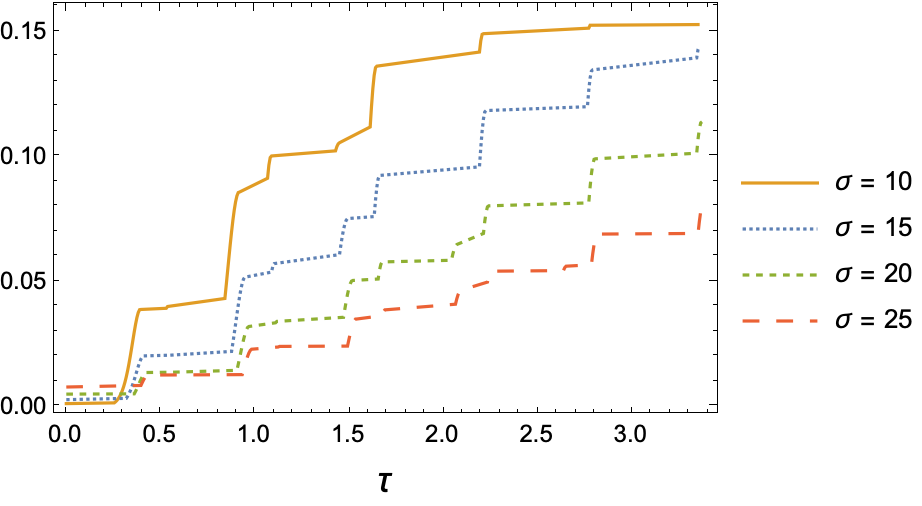}\caption{Maximum variance over time for the $d=200$, $c=1.05$ quantum three-wave
interaction. Results for various initial standard deviations are shown.}
\label{fig:variance}
\end{figure}

So far, the initial standard deviation of the quantum system has been
taken to be $\sigma=d/5$ for simplicity; however, this does not necessarily
lead to the best correspondence with the classical system. There is
a trade off between long-term fidelity and the size of the initial
variance. Shown in Fig. \ref{fig:variance} is the growth of the variance
over time given different initial variances. Only the maximum variance
is plotted because the magnitude of the variance oscillates with the
amplitude of the wave. This can be seen in the spurts of growth of
the maximum variance occurring at multiples of the period of the nonlinear
oscillation. For a higher initial standard deviation, the variance
grows more slowly with time. 

The slow growth of the variance with larger initial standard deviations
is attributable to the spectrum of the Hamiltonian, shown in Fig.
\ref{fig:variance}. When $c\rightarrow\infty$, the $d$ eigenvalues
will be exactly linearly distributed, and for finite $c>1$, they
will lie between the two lines of the figure. Importantly, even for
the maximally nonlinear $c=1$, the eigenvalues with small absolute
value (those with scale eigenvalue indices of around $0.5$) will
still be approximately linearly distributed. As the three-wave interaction
acts similarly to a perturbed quantum harmonic oscillator, we may
by analogy understand that states with a large initial standard deviation
are represented by the lowest energy eigenmodes---those eigenmodes
which reside in the central, linear area of Fig. \ref{fig:variance}.
With higher initial variance states having their dominant eigenfrequencies
linearly distributed, they remain in correspondence with the classical
dynamics for longer periods of time. Thus, despite a high initial
variance coming at the cost of the quantum Eq. (\ref{eq:scaled_quantum})
beginning with dynamics farther from those of the classical Eq. (\ref{eq:scaled_classical_eqn})
with no variance, the effect of quantum interference is suppressed.
Of course, if the initial variance is taken to be too large, 1) the
initial condition may no longer be taken to be Gaussian, and 2) the
initial condition ceases to be well-approximated by lower frequency
eigenstates. An initial standard deviation of $\sigma=d/5$ strikes
a balance between the need for a small growth rate of the variance
with the need to prevent clipping of the initial Gaussian in the finite
domain with $\alpha_{-1}=\alpha_{d}=0$ enforced. 
\begin{figure}
\includegraphics[scale=0.75]{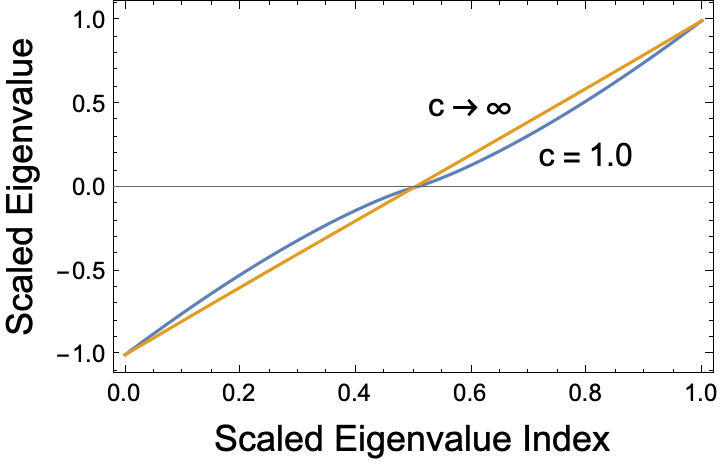}\caption{Eigenvalues of the Hamiltonian scaled by the highest eigenvalue. The
indices of the eigenvalues linearly increase from that of the lowest
eigenvalue 0, to the highest index 1. For finite dimensional systems,
the eigenvalues will lie between the $c=1$ and $c\rightarrow\infty$
curves. }
\label{fig:spectrum}
\end{figure}

\begin{figure}
\noindent\begin{minipage}[t]{1\columnwidth}%
\includegraphics[width=0.4\linewidth]{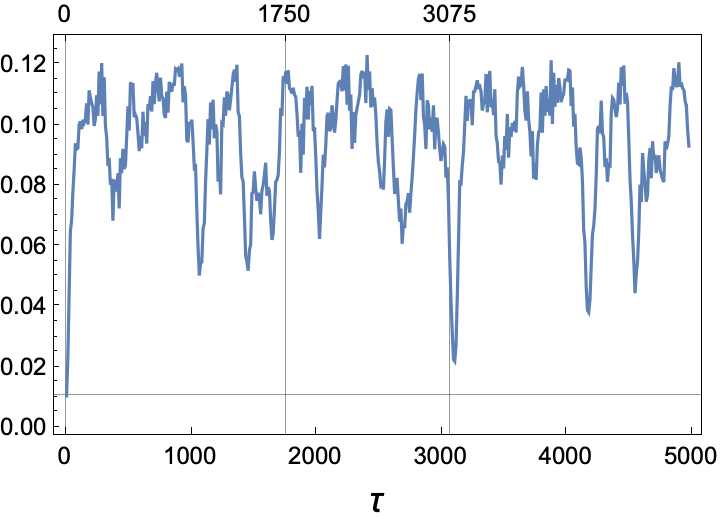}%
\end{minipage}\caption{Variance of $d=100$ quantum system with nonlinearity parameter $c=1.05$
and initial standard deviation set to $\sigma=d/5=20$. The three
labelled vertical gridlines at $\tau=0$, $1750$, and $3075$ indicate
the starting times of the three elements of Fig. \ref{fig:Quantum_Revival}.
The horizontal gridline shows the starting variance $\delta^{\prime}=0.011$.
\label{fig:variance_for_revival}}
\end{figure}

\begin{figure}
\noindent\begin{minipage}[t]{1\columnwidth}%
\includegraphics[width=1\linewidth]{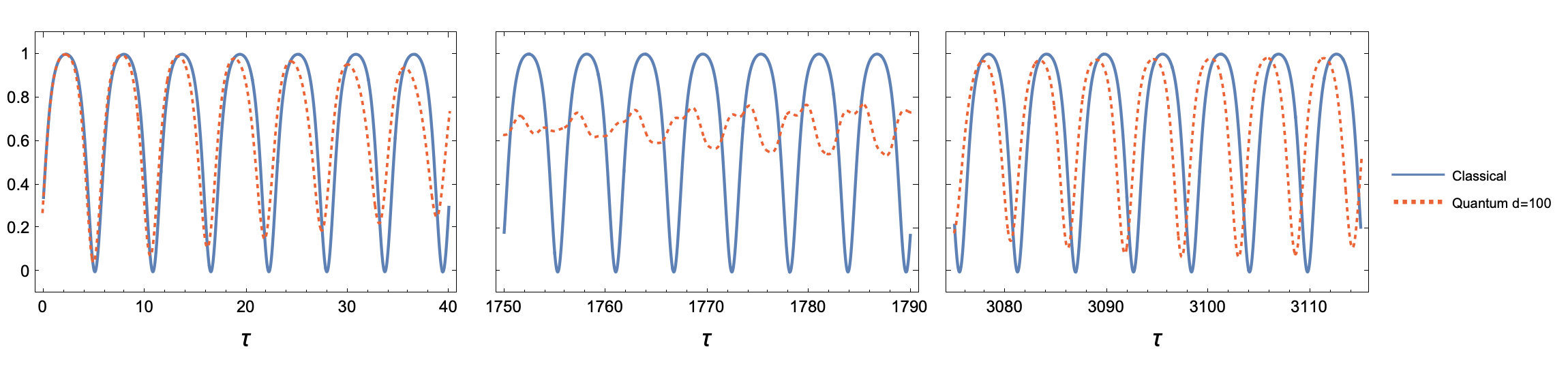}%
\end{minipage}\caption{Comparison between classical and quantum dynamics for nonlinearity
parameter $c=1.05$ beginning at three times $\tau=0$, $1750$, and
$3075$. The plot beginning at $\tau=1750$ typifies high variance
behavior, and the plot beginning at time $\tau=3075$ shows a partial
quantum revival, which was found by looking for relative minima in
the variance of Fig. \ref{fig:variance_for_revival}. The $d=100$
quantum system begins with a standard deviation of $\sigma=d/5=20$.
\label{fig:Quantum_Revival} }
\end{figure}

The variance may be used as a tool to determine where quantum revivals
will occur. Shown in Fig. \ref{fig:variance_for_revival} is the variance
for the $d=100$ quantum system for the first approximately 880 classical
orbits. The initial standard deviation is $\sigma=20$, which corresponds
to an initial variance of $\delta^{\prime}=0.011$. Beginning around
$\tau=3075$, the variance sharply decreases to a minimum of $\delta^{\prime}=0.022$
indicating a partial quantum revival, shown in Fig. \ref{fig:Quantum_Revival}.
Although the phase and amplitude of the quantum system differ from
that of the classical, during the revival, the quantum system accurately
captures the period of the classical orbit. When the variance is higher,
for example, beginning at $\tau=1750$, it is difficult to characterize
the correlation between the quantum and classical systems. 

\section{Conclusions\label{sec:Conclusions}}

As a model for the classical system, the quantum description of the
three-wave interaction has several advantages. First, via the correspondence
principle, the classical and quantum systems are guaranteed to converge
as the dimension of the quantum system is increased. Second, we did
not need to rely on closures or arbitrary choices of a finite representation,
as would have been the case for a KvN quantization of the classical
dynamics. The quantum field-theoretic method for quantization provides
a systematic means of rendering the dynamics discrete, finite-dimensional,
and unitary. Finally, as displayed in Figs. \ref{fig:quantum_classical_correspondence}
and \ref{fig:error}, the quantum system of appropriate dimension
can capture both the qualitative and quantitative aspects of the nonlinear
periodic solution, including its frequency. 

While the dimension of the quantum systems being compared to classical
dynamics have been large ($d\sim100$), simulations of this degree
may soon be achievable with quantum computers. Since $n\propto log_{2}(d)$,
where $n$ is the number of qubits necessary to represent a state,
neglecting error correction, a $d=100$ quantum state only requires
8 qubits to represent it. However, difficulties would arise from approximating
the three-wave Hamiltonian as a series of universal gates acting on
those qubits. In general, approximating arbitrary unitary operators
can require $O(d^{2})$ gate applications, which is obviously untenable
for high dimension. Shi \textit{et al.}\textit{\emph{, in simulations
of the quantum three-wave equations with $d=3$, have shown this problem
may be sidestepped, though, through the creation of special-made gates
particular to the system one is simulating \citep{shiPhysRevA}. }}

Another hurdle to quantum simulation of the three-wave interaction
relates to the information being extracted from the quantum system.
Measuring the full time history of each component of the quantum phase
space, $\alpha_{i}$, will destroy any potential quantum speedup;
however, when compared with simulating the full classical Liouville
dynamics, simulating the quantum dynamics may result in speedups as
long as the extracted information is sparse. In particular, we have
shown that low-dimensional classical information, including the nonlinear
frequency and the expectation value of the number operator, may be
effectively simulated in a quantum system. \textit{\emph{While the
three-wave interaction is only the lowest order nonlinearity in plasma
physics, this opens the possibility of using natively unitary quantum
dynamics to model more complicated classical, nonlinear dynamics on
quantum hardware in the near future. }}
\begin{acknowledgments}
This research was supported by the U.S. Department of Energy (DE-AC02-09CH11466).
\end{acknowledgments}

\bibliographystyle{apsrev4-2}
\bibliography{draft}

\end{document}